\documentclass[twocolumn,prl,aps,superscriptaddress,showpacs]{revtex4}
\usepackage{epsfig}

\begin{document}

\title{Fermi Arcs in a Doped Pseudospin-1/2 Heisenberg Antiferromagnet}

\author{Y. K.  Kim}
\affiliation{Advanced Light Source, Lawrence Berkeley National Laboratory, Berkeley, CA 94720, USA}

\author{O. Krupin}
\affiliation{Advanced Light Source, Lawrence Berkeley National Laboratory, Berkeley, CA 94720, USA}

\author{J. D. Denlinger}
\affiliation{Advanced Light Source, Lawrence Berkeley National Laboratory, Berkeley, CA 94720, USA}

\author{A. Bostwick}
\affiliation{Advanced Light Source, Lawrence Berkeley National Laboratory, Berkeley, CA 94720, USA}

\author{E. Rotenberg}
\affiliation{Advanced Light Source, Lawrence Berkeley National Laboratory, Berkeley, CA 94720, USA}

\author{Q. Zhao}
\affiliation{Materials Science Division, Argonne National Laboratory, Argonne, IL 60439, USA}

\author{J. F. Mitchell}
\affiliation{Materials Science Division, Argonne National Laboratory, Argonne, IL 60439, USA}

\author{J. W. Allen}
\affiliation{Randall Laboratory of Physics, University of Michigan, Ann Arbor, MI 48109, USA}

\author{B. J. Kim}
\email[Electronic address:$~~$]{bjkim@fkf.mpg.de}
\affiliation{Materials Science Division, Argonne National Laboratory, Argonne, IL 60439, USA}
\affiliation{Randall Laboratory of Physics, University of Michigan, Ann Arbor, MI 48109, USA}
\affiliation{Max Planck Institute for Solid State Research, Heisenbergstra§e 1, D-70569 Stuttgart, Germany}

\date{\today}

\begin{abstract}
High temperature superconductivity in cuprates arises from an electronic state that remains poorly understood. We report the observation of a related electronic state in a non-cuprate material Sr$_2$IrO$_4$ in which the unique cuprate Fermiology is largely reproduced. Upon surface electron doping through in situ deposition of alkali-metal atoms, angle-resolved photoemission spectra of Sr$_2$IrO$_4$ display disconnected segments of zero-energy states, known as ÔFermi arcsÕ, and a gap as large as 80 meV. Its evolution toward a normal metal phase with a closed Fermi surface as a function of doping and temperature parallels that in the cuprates. Our result suggests that Sr$_2$IrO$_4$ is a useful model system for comparison to the cuprates. 

\end{abstract}
\maketitle


%

Although the mechanism of high temperature superconductivity (HTSC) remains an open question, it is commonly believed that certain unique features of cuprates are essential to high-temperature superconductivity: spin-1/2 moment on a quasi-two-dimensional square lattice, Heisenberg antiferromagnetic coupling, and no orbital degeneracy. A minimal model based on this assumption can reproduce much of the phenomenology of the cuprates\cite{1}. Within this framework, it would be informative to realize the key features of cuprates in a different material\cite{2}. The 5$d$ transition-metal oxide Sr$_2$IrO$_4$ with a ${t_{2g}}^5$ valence shell is a Mott insulator in which the orbital degeneracy is removed through strong spin-orbit coupling\cite{3,4}. Despite strong entanglement of spin and orbital degrees of freedom, the resulting pseudospins (with $J_{\text eff}$=1/2 quantum number) exhibit the spin dynamics of a Heisenberg antiferromagnet\cite{5} with the nearest-neighbor magnetic exchange coupling having an energy scale on the order of 60-100 meV\cite{6,7}. As a result, despite the very different starting electronic structures, the effective low-energy physics of Sr$_2$IrO$_4$ can be described by the same minimal model as developed for the cuprates, with comparable magnitudes of coupling constants, suggesting possible unconventional HTSC upon carrier doping of this material\cite{8}; indeed, a recent Monte Carlo study predicts a $d$-wave superconducting phase for the electron-doped case\cite{9}.

\begin{figure}
\centering \epsfxsize=7 cm \epsfbox{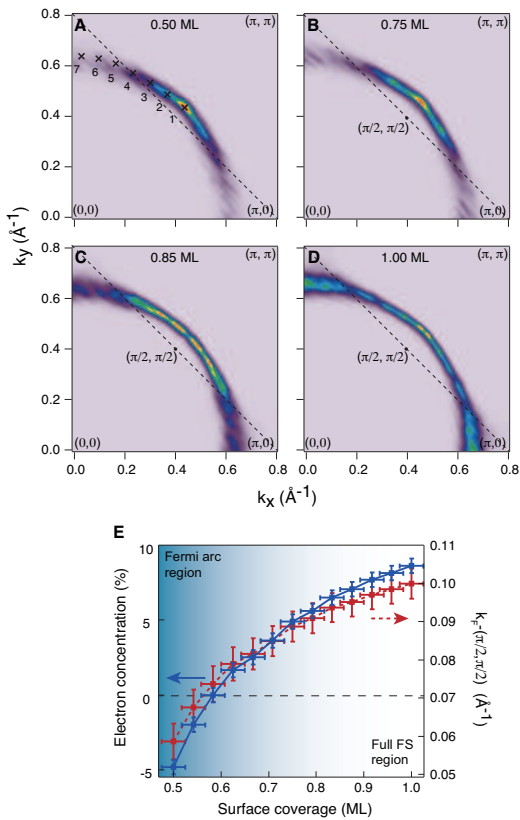} \caption{(Color online) Constant-energy intensity maps taken at $E_{\text F}$ for surface coverage of (A) 0.5 ML, (B) 0.75 ML, (C) 0.85 ML, and (D) 1 ML, shown over a quadrant of the Brillouin zone of the undistorted square lattice. The actual Brillouin zone respecting the $\sqrt{2}\times\sqrt{2}$ superstructure of the lattice distortion coincides with the magnetic Brillouin zone whose boundary is indicated by the dashed line. The intensity maps were normalized by the angular profile of the intensity along the FS measured at 1 ML. The map was symmetrized with respect to the diagonal line connecting (0,0) and ($\pi$,$\pi$). (E) The electron count based on the volume enclosed by the apparent large Fermi surface centered at the $\Gamma$ point and the distance from the node to ($\pi$/2,$\pi$/2) as a function of coverage. Zero of the left vertical axis corresponds to the case of half-filled band. }
\end{figure}

However, little is so far known experimentally about the nature of the electronic phases that derive from carrier doping of Sr$_2$IrO$_4$\cite{10,11}. One of the outstanding questions is whether the analogy to cuprates can be extended to the case of a metallic phase, because Sr$_2$IrO$_4$ may evolve differently with doping. In particular, weakening of spin-orbit coupling may cause the system to revert to a multi-band metal.  On the other hand, if the single-band picture remains intact, one may expect the generic cuprate phenomenology to be reproduced in this material. To address this issue, we use an in situ surface doping technique, an approach that has been proven effective for surface doping control of complex oxides\cite{12}, to reveal the doping evolution of the electronic structures using angle-resolved photoemission. By varying the surface coverage of potassium atoms deposited on the surface of the parent insulator Sr$_2$IrO$_4$, we follow the electronic evolution across the entire phase diagram from a Mott insulating to a normal metallic state, via a ÔstrangeÕ metallic phase that bears marked spectral similarities to those observed for the cuprates. We present our data in units of surface coverage rather than in electron doping concentration because of a highly nonlinear relation between the two.

Figures 1A-1D show the constant energy intensity maps at the Fermi level ($E_{\text F}$) at T=70 K for surface coverage ranging from 0.5 monolayer (ML) to 1 ML. At 1 ML, we observe a closed, large Fermi surface (FS) indicative of a normal metal phase with the enclosed area of 54.3 $\%$ of the two-dimensional Brillouin zone (BZ) corresponding to 8.6 $\%$ electron doping relative to a half filled band. With decreasing coverage, however, we observe two gradual but conspicuous spectral changes that indicate formation of a distinct electronic phase. First, the intensity is much suppressed in an extended region near ($\pi$,$\pi$), breaking up the large FS into disconnected segments of Fermi arcs\cite{13,14}. [Hereafter we refer to the maximum (minimum) intensity point as ÔnodeÕ (ÔantinodeÕ) in analogy to cuprates]. Second, the locus of the node shows a large shift toward ($\pi/2$,$\pi/2$) (Fig. 1E). As a result, an attempt to interpret the arc as a part of the large FS encounters a qualitative inconsistency in the electron count based on LuttingerÕs sum rule: the volume enclosed by the apparent large FS becomes less than half of the BZ for coverage below Å0.6 ML (Fig. 1E), in clear contradiction with electron doping. These results imply either a global change in the FS topology or an inconsistency with the notion of FS as the system departs from the normal metal phase observed at 1 ML.  We note that an analogous qualitative inconsistency in the hole count occurs\cite{15} when interpreting the arcs in cuprates\cite{16} as part of an underlying large FS Ð an initial hint that the analogy to cuprates does indeed extend to the case of a metallic phase.

\begin{figure*}
\centering \epsfxsize=16.5 cm \epsfbox{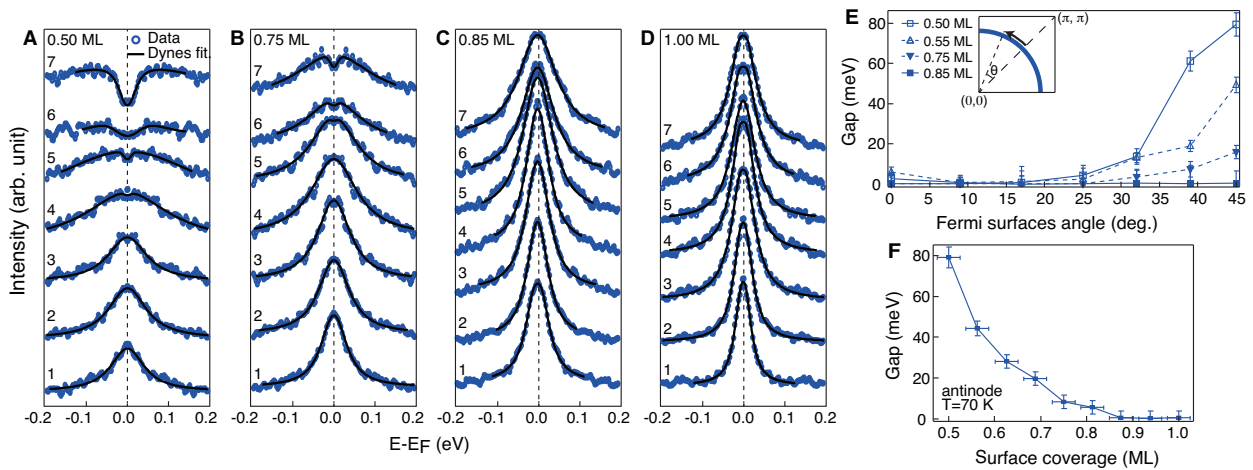} \caption{(Color online) Symmetrized EDCs (blue open circles) fit to the Dynes formula (black solid lines) along the arc at 7 $k_{\text F}$ points labeled in Fig. 1A for surface coverage of (A) 0.5 ML, (B) 0.75 ML, (C) 0.85 ML, and (D) 1 ML. (E) The gap magnitude $\Delta$ as a function of Fermi surface angle ($\theta$) defined in the inset. (F) The gap magnitude $\Delta$ at the antinode as a function of surface coverage. .}
\end{figure*}

The ÔFermi arcÕ metal phase can be characterized by an order parameter that depends on both the coverage and temperature. Figures 2A-2D show the energy distribution curves (EDCs) for T=70 K, symmetrized to remove the effect of the Fermi function. In the antinodal region, the symmetrized EDCs show a clear opening of a gap at 0.5 and 0.75 ML, which closes as the system enters the normal metal state at higher coverage. We use a phenomenological Dynes formula\cite{17}, widely used to fit superconducting gaps\cite{18,19}, to quantify the gap. Thereby we find for T=70 K and 0.5 ML a gap as large as Å80 meV, which rapidly decreases away from the antinode and becomes zero over a finite length of the contour that defines an arc (Fig. 2E). The gap at the antinode continuously diminishes with increasing surface coverage and becomes undetectable within the energy resolution (Å18 meV) beyond 0.85 ML (Fig. 2F).

Figures 3A-3C show the $E_{\text F}$ intensity map taken at T=30 K, 70 K, and 110 K for surface coverage of 0.7 ML. Similar to the surface coverage dependence, with increasing temperature the arc expands its enclosed volume (when extrapolated to a large FS), crossing from less than half-filling at T=30 K (48.9 $\%$) to more than half-filling (52.1 $\%$ of BZ) at T=110 K. Concomitantly, the arc elongates (Fig. 3F) with increasing temperature and evolves to a closed FS at T=110 K. On cooling down to T=30 K, the length of the arc shortens considerably but a finite length of arc still remains (Fig. 3F), which is incompatible with a $d$-wave gap indicative of superconductivity. Unfortunately, measurement at lower temperatures to check for possible emergence of a $d$-wave gap was precluded by the high resistance of the sample, the bulk of which remains an insulator.

Consistent with the formation of a closed FS at high temperature, the antinode gap magnitude decreases with increasing temperature as indicated by the gradual shift of the leading edge in the EDC (Fig. 3E, 3F inset), in clear contrast to the temperature evolution at the node that shows only a minor thermal broadening (Fig. 3D). The gap starts to emerge below T=110 K but grows slowly with decreasing temperature, in contrast to the typical behavior of an order parameter that sets in at a well-defined transition temperature. However, the significant temperature dependence indicates a non-trivial origin and a many-body nature of the gap. We remark that the overall momentum\cite{19} and temperature\cite{20,21} dependences of the gap closely follow those observed for the pseudogap in cuprates.

\begin{figure}
\centering \epsfxsize=8.5 cm \epsfbox{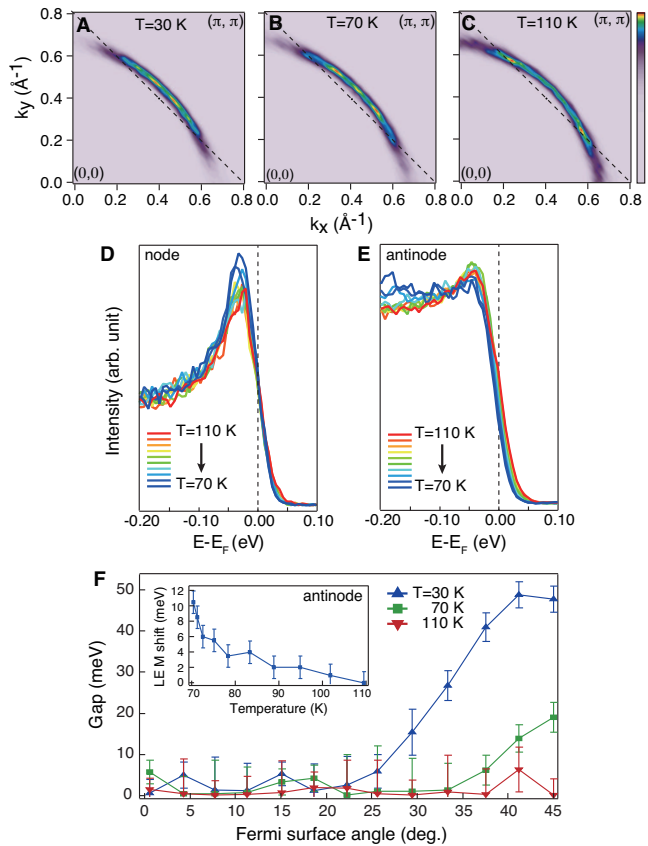} \caption{(Color online) $E_{\textrm{F}}$ intensity map for (A) T=30 K, (B) T=70 K, and (C) T=110 K. EDCs at the (D) node and (E) antinode in temperature range between 70 and 110 K. (F) The gap magnitude $\Delta$ as a function of Fermi surface angle for T=30 K, 70 K, and 110 K. Inset shows the temperature dependence of the antinode leading edge midpoint shift, which tracks approximately one-half of the antinode gap magnitude\cite{32,33}.}
\end{figure}

All the preceding results Ð carrier counting inconsistency and gap behavior Ð indicate that the parallel between Sr$_2$IrO$_4$ and cuprates persists in the metallic phase induced by electron doping, and that the Ôbreaking upÕ of a Fermi surface into disconnected segments is a general phenomenon of a system containing certain generic characteristics of cuprates. The formation of an arc, as opposed to Ôhot spotsÕ at intersections with magnetic zone boundary, indicates that the correlation strength in Sr$_2$IrO$_4$, which has been under intense debates\cite{22,23,24,25,26}, is in the intermediate-to-strong coupling regime\cite{27}. Thus, Mott physics and local correlations are essential for understanding of the physics of Sr$_2$IrO$_4$. Taking all these together, we conclude that the phenomenology of the Fermi arc must be accountable within the minimal description of the physics of doping a single band, spin-1/2 antiferromagnetic Mott insulator, as shared by Sr$_2$IrO$_4$ and cuprates. Their differences in microscopic electronic structures help separate essential features of HTSC from material-specific features. In particular, whether the charge gap is of Mott insulating (Sr$_2$IrO$_4$) or of charge transfer nature (cuprates) seems to have little bearing on the properties discussed here. 

The nature of the electronic phase manifesting Fermi arcs in electron-doped Sr$_2$IrO$_4$ remains unclear at this point. A growing body of evidence in cuprates that Fermi arcs are associated with distinct phases that compete with HTSC, such as density waves\cite{28}, stripes\cite{29,30}, and checkerboard orders\cite{31}, prompts investigation of competing phases in carrier-doped Sr$_2$IrO$_4$. This in turn will help clarify which of these phases, if any, occur generically in proximity to HTSC. If, on the other hand, Fermi arcs are a precursor signature of $d$-wave superconductivity, Sr$_2$IrO$_4$ may superconduct at lower temperatures. Further inquiries into these questions will shed new light on the long-sought connection between the pseudogap and HTSC.

\acknowledgments
We acknowledge helpful discussions with J. H. Shim, K. Haule, G. Kotliar. C. Kim, M. Norman, and G. Khaliullin. This work was supported at UM by the US National Science Foundation under grant number DMR-07-04480. B.J.K. acknowledges the Institute for Complex Adaptive Matter for a travel grant that enabled a visit and helpful discussions at Rutgers University. Work in the Materials Science Division of Argonne National Laboratory (sample preparation and characterization) was supported by the U.S. Department of Energy Office of Science, Basic Energy Sciences, Materials Science and Engineering Division.  The Advanced Light Source is supported by the Director, Office of Science, Office of Basic Energy Sciences, of the U.S. Department of Energy under Contract No. DE-AC02-05CH11231. Y. K. Kim is supported through NRF Grants funded by the MEST (No. 20100018092).

\setcounter{figure}{0}
\renewcommand{\thefigure}{S\arabic{figure}}

\section{supplemental material}

\subsection{Calibration of the growth rate}

The method of in situ surface doping has been successfully applied to some cuprates\cite{12,34} and iridates\cite{35}, but the growth mode and rate can vary from one compound to another. We calibrated the growth rate in Sr$_2$IrO$_4$ by monitoring the surface coverage through the quantum well states formed in the potassium (K) overlayer as well as the K 3$p$ core level spectrum. Fig. S1A shows the free-electron-like quantum well bands (yellow dotted lines) from the K overlayer superimposed on the spectral features from the doped sample surface. Note that the quantum well band minima align with the sample $\Gamma$ point, implying that the K overlayer is growing epitaxially. The quantum well bands appear in every other Brilloiun zone, or equivalently, follow the ÒoriginalÓ Brillouin zone without the $\sqrt{2}\times\sqrt{2}$ lattice distortion due to octahedral rotation. 
\begin{figure}[!hb]
\centering \epsfxsize=8.5cm \epsfbox{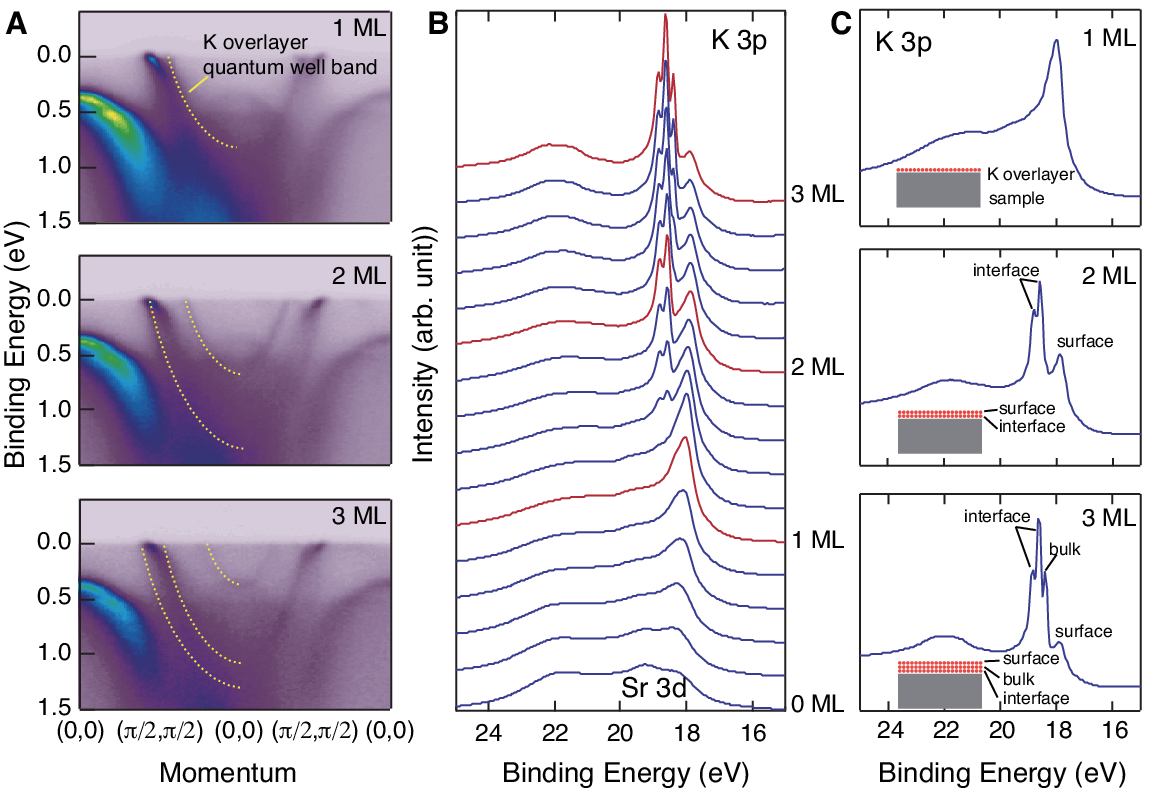} \caption{(Color online) (A) ARPES spectra along (0,0)-($\pi/2$,$\pi/2$)-(0,0) direction at three different thicknesses (1, 2, and 3 monolayers) of K overlayer.  Yellow dotted lines show the quantum well bands. (B) K 3$p$ core level spectrum as a function of K overlayer thickness. (C) Chemically shifted K 3$p$ core level peaks assigned as from surface, bulk, and interface, following Ref.\cite{36}}
\end{figure}

These quantum well bands are most intense when the number of layers grown is close to an integer, implying the layer-by-layer growth mode. The thickness of the overlayer at integer values can be read off from the number of the quantum well bands, which correlates well with the thickness inferred from the core level spectra shown in Fig. S1B. The bottom curve in Fig. S1B is from the as-cleaved sample surface, which shows the Sr 3$d$ core level peaks that overlap in energy with the K 3$p$ core level. The K 3$p$ core level peak grows in intensity as K atoms are deposited on the sample surface. As the second K overlayer starts to grow, a pair of chemically shifted K 3$p$ peaks appears at a higher binding energy. Likewise, a third peak appears when the third layer starts to grow. The evolution of the K 3$p$ core level peaks as a function of thickness is well documented in Ref.\cite{36}, which is reproduced in Fig. S1C.

\subsection{Characterization of samples grown from K$_2$CO$_3$ flux} 

The pristine samples grown in SrCl$_2$ flux charged below T=70 K, which precluded ARPES measurement at lower temperatures. We grew Sr$_2$IrO$_4$ in K$_2$CO$_3$ flux to deliberately add K impurities, which resulted in a resistivity several orders of magnitude lower than that of pristine samples and allowed ARPES measurement at temperatures as low as T=30 K. The samples were phase pure as checked by powder x-ray diffraction (not shown). While the concentration of K was below our detection limit of energy dispersive x-ray spectroscopy, the significantly lower onset temperature (T$\sim$200 K) of the large in-plane magnetic response indicated finite doping of K, presumably replacing Sr ions. 
  
\begin{figure}
\centering \epsfxsize=8.5cm \epsfbox{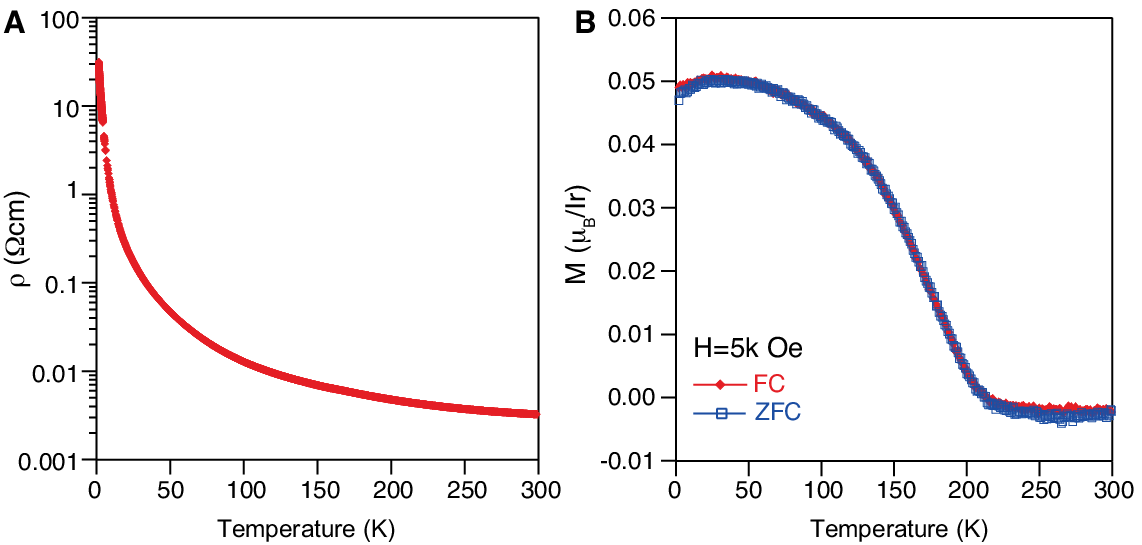} \caption{(Color online) (A) Resistivity and (B) magnetization of a representative sample grown from K$_2$CO$_3$ flux.}
\end{figure}
 \subsection{Material and method}
ARPES measurements were performed at Beamline 7.0.1 and Beamline 4.0.3 at the Advanced Light Source, equipped with Scienta R4000 electron analyzers. We evaporated potassium atoms in situ onto the freshly cleaved surface of Sr$_2$IrO$_4$ using a commercial SAES evaporator to dope electrons into the sample surface. The samples, grown from SrCl$_2$ flux, were cleaved, doped, and measured at T=70 K and p$\sim$3$\times$10$^{-11}$ Torr with an overall energy resolution of 18 meV. The data were reproduced in more than 5 samples. In order to perform measurement at lower temperatures, we also grew samples using different flux materials to deliberately introduce impurities to reduce charging. Using samples grown from K$_2$CO$_3$ flux, we were able to measure at a temperature as low as 30 K.


\begin{thebibliography}{30}
\bibitem{13} Note that the positive sign of next-nearest hopping tÕ renders a FS topology with a large electron pocket around the $\Gamma$ point, as opposed to a hole pocket centered at ($\pi$,$\pi$) in cuprates with a negative tÕ (8).
\end{thebibliography}
\end{document}